\documentclass[preprint,superscriptaddress,preprintnumbers,amsmath,amssymb,pra]{revtex4-1}

\usepackage{graphicx}% Include figure files
\usepackage{dcolumn}% Align table columns on decimal point
\usepackage{bm}% bold math
\usepackage{CJK}
\usepackage{appendix}
\font\scripti=cmmi7
\font\scriptscripti=cmmi5
\def\sib#1{\setbox0 = \hbox{\scripti #1}
  \kern-.02em\copy0\kern-\wd0
  \kern.04em\box0} % script italic bold 
\def\ssib#1{\setbox0 = \hbox{\scriptscripti #1}
  \kern-.02em\copy0\kern-\wd0
  \kern.04em\box0} % scriptscript italic bold
\font\tenib=cmmib10 % italic bold for math
\skewchar\tenib='177 \skewchar\tenib='177 \skewchar\tenib='177
\def\pbold#1{\setbox0 = \hbox{$ #1 $}
  \kern-.022em\copy0\kern-\wd0
  \kern.011em\copy0\kern-\wd0
  \kern.011em\copy0\kern-\wd0
  \kern.011em\copy0\kern-\wd0
  \kern.011em\box0} % poorman's bold
\usepackage{graphicx}% Include figure files
\usepackage{dcolumn}% Align table columns on decimal point
\usepackage{bm}% bold math
\usepackage{color}

\def\lesssim{\ \raise.3ex\hbox{$<$}\kern-0.8em\lower.7ex\hbox{$\sim$}\ }
\def\gesim{\ \raise.3ex\hbox{$>$}\kern-0.8em\lower.7ex\hbox{$\sim$}\ }

\newcommand{\red}[1]{{{#1}}}

\begin{document}
%\begin{CJK}{UTF8}{ipxm}
\preprint{RIKEN-iTHEMS-Report-22}

\title{Role of the effective range in the density-induced BEC-BCS crossover}

\author{Hiroyuki Tajima}
\email{hiroyuki.tajima@tnp.phys.s.u-tokyo.ac.jp}
\affiliation{Department of Physics, Graduate School of Science, The University of Tokyo, Tokyo 113-0033, Japan}

\author{Haozhao Liang}
\email{haozhao.liang@phys.s.u-tokyo.ac.jp}
\affiliation{Department of Physics, Graduate School of Science, The University of Tokyo, Tokyo 113-0033, Japan}
\affiliation{RIKEN iTHEMS, Wako 351-0198, Japan}

\date{\today}
\begin{abstract}
We elucidate the role of the effective range in the Bose-Einstein-condensate (BEC) to Bardeen-Cooper-Schrieffer (BCS) crossover regime of two-component fermions \red{in three dimensions}.
In contrast to ultracold Fermi gases near the broad Feshbach resonance, where the interaction can be characterized by the contact-type interaction,  
the interaction range in general becomes important in the density-induced BEC-BCS crossover discussed in the context of condensed-matter and nuclear systems. 
Characterizing the non-local interaction in terms of the low-energy constants such as scattering length $a$ and effective range $r$, we show how the crossover phenomena are affected by nonzero effective ranges.
In particular, we show that the superfluid order parameter is strongly suppressed in the high-density regime and the sound velocity exhibits a non-monotonic behavior reflecting mechanical stability of the system.
Moreover, we point out that the high-momentum tail associated with the contact parameter can be visible when the magnitude of momentum $k$ is much less than $1/r$.

\end{abstract}
%\pacs{03.75.Ss, 03.75.-b, 03.70.+k}
\maketitle
%%%%%%%%%%%%%%%%%%%%%%%%%%%%%%%%%%%%%%%%%%%%%%%%%%%%%%%%%%%%%%%%%%%%%%%%%%%%%
%\par
\section{Introduction}

The realization of the crossover from the Bose-Einstein condensate (BEC) to Bardeen-Cooper-Schrieffer (BCS) pairing in ultracold Fermi gases~\cite{Regal2004PhysRevLett.92.040403,Zwierlein2004PhysRevLett.92.120403,Bartenstein2004PhysRevLett.92.203201}
developed a new frontier for exploring strongly interacting quantum systems.
Recently, such a many-body phenomenon called BEC-BCS crossover, which was originally studied by Eagles~\cite{Eagles1969PhysRev.186.456} and Leggett~\cite{leggett1980modern}, 
has been extensively studied~\cite{CHEN20051,BCS-BEC,STRINATI20181,OHASHI2020103739}. 

The concept of the BEC-BCS crossover has also been discussed in the various systems such as neutron matter~\cite{Gandolfi2015annurev-nucl-102014-021957,ramanan2021pairing}
and dense quark matter~\cite{Abuki2002PhysRevD.65.074014,Nishida2005PhysRevD.72.096004,He2007PhysRevD.75.096003},
and moreover nowadays such crossover phenomena have been discovered in unconventional superconductors~\cite{Kasahara2014PNAS,nakagawa2021gate,Suzuki2022PhysRevX.12.011016} as well as in electron-hole systems~\cite{liu2022crossover}. 
On the other hand, the mechanism for driving the system to the strongly-interacting regime by changing the density and the pressure is different from the BEC-BCS crossover in an ultracold Fermi gas where the magnetic Feshbach resonance is used to enhance the scattering length $a$~\cite{Chin2010RevModPhys.82.1225}.

On the basis of the scattering theory,
the interaction strength can be measured by using the dimensionless coupling parameter $1/(k_{\rm F}a)$ where $k_{\rm F}$ is the Fermi momentum.
\begin{figure}[t]
    \centering
    \includegraphics[width=8cm]{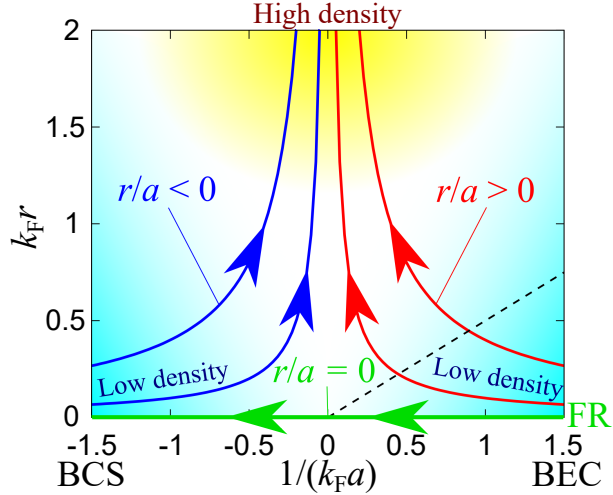}
    \caption{Evolution of coupling parameters in the BEC-BCS crossover in terms of the scattering length $a$ and the effective range $r$ $(>0)$, normalized by the Fermi momentum $k_{\rm F}$.
    The dashed line shows $\red{\cot}\delta_s(k=k_{\rm F})=0$~\cite{Tajima2019JPSJ}.
    While ultracold Fermi gases near the broad Feshbach resonance (FR) undergo the crossover by tuning $a$ with small $r$ (i.e., $r/a\simeq 0$), the density-induced  crossover discussed in other condensed-matter and nuclear systems runs on the lines with fixed ratio $r/a$. 
    If the two-body bound state exists in the dilute limit ($r/a>0$) such as excitons in electron-hole systems, deuterons in nuclear matter, and baryons in two-color QCD, the system evolves from the BEC regime to the large effective-range regime with increasing the density $\rho\propto k_{\rm F}^3$.
    On the other hand, in the absence of two-body bound state ($r/a<0$), the systems start from the BCS regime to the large effective-range regime.
    }
    \label{fig:1}
\end{figure}
Therefore, the BEC-BCS crossover is realized by tuning $a$ from positive values to negative ones in cold atomic systems as shown in Fig.~\ref{fig:1}.
On the other hand, in the case of the BEC-BCS crossover realized in other systems \red{by changing $k_{\rm F}$ with fixed interaction parameters such as $a$ (in the following, we call it as the {\it density-induced} BEC-BCS crossover)},
one can consider that $1/|k_{\rm F}a|$ becomes smaller with increasing the density.
Hence, if the two-body bound state exists in the dilute limit ($a>0$),
the system undergoes the crossover from the BEC regime [$1/(k_{\rm F}a)\gesim 1$] to the unitary regime [$1/(k_{\rm F}a)\simeq 0$] with increasing $k_{\rm F}$.
However, in the high-density regime, the short-range properties of the interaction cannot be negligible in contrast to ultracold Fermi gases near the broad Feshbach resonance~\cite{BCS-BEC}, where the interaction range is much less than the other length scales such as $a$.
In such a case, the leading-order contribution can be an effective-range corrections associated with the low-momentum expansion of the $s$-wave phase shift $\red{k\cot}\delta_{s}(k)=-\frac{1}{a}+\frac{1}{2}rk^2+O(k^4)$,
where $r$ is called effective range.
Indeed, the dimensionless parameter $k_{\rm F}r$ characterizing the importance of $r$ also increases with $k_{\rm F}$.
In this regard, the effective range corrections can be significant in the high-density regime of the density-induced BEC-BCS crossover.

The effective range corrections based on the $k_{\rm F}r$ expansion have been investigated in terms of the similarity between neutron matter and ultracold Fermi gases~\red{\cite{Schwenk2005PhysRevLett.95.160401,Forbes2012PhysRevA.86.053603,Schonenberg2017PhysRevA.95.013633,PhysRevC.77.032801}}.
Moreover, in the high-density regime (corresponding to the regime with large $k_{\rm F}r$), the occurrence of the mechanical collapse
has also been reported~\cite{Schonenberg2017PhysRevA.95.013633}, being similar to electron-hole droplet in semiconductor systems~\cite{keldysh1986electron} as well as liquid-gas transition in nuclear matter~\cite{Jin2010PhysRevC.82.024911}.
The cluster formation has also been discussed in a few-body system with finite-range interactions~\cite{Yin2019PhysRevLett.123.073401}.
Moreover, one may expect that the mechanical property is related to the non-monotonic behavior of the sound velocity extensively discussed in the hadron-quark crossover of massive stars~\cite{kojo2021qcd} because the similar crossover has been discussed in two-color quantum chromodynamics (QCD)~\cite{iida2020two,Kojo2022PhysRevD.105.076001}.
Although in general the high-density regime can be affected by the short-range sector of the interaction including the effective range,
it is worth investigating how such a mechanical collapse can occur with the interaction potential with the {\it pure} effective-range corrections without any higher order coefficients~\cite{Wyk2018PhysRevA.97.013601,Tajima2019JPSJ}.
Such a study can be helpful for further understanding of generalized universal properties of many-body fermions with non-local interactions, stepping further from the contact-type interaction characterized by only $a$.
In addition, the optical control of the effective range has been proposed in cold atomic systems~\cite{Wu2012PhysRevLett.108.010401,Wu2012PhysRevA.86.063625} and related experiments are ongoing~\cite{Arunkumar2018PhysRevLett.121.163404,Arunkumar2019PhysRevLett.122.040405}.
In this regard, the system with large effective range and scattering length but sufficiently small higher-order coefficients in the phase shift can be realized in the future experiments.
The equation of state in such a system can also be precisely measured by using the recent state-of-the-art imaging technique~\cite{Horikoshi2017PhysRevX.7.041004,Tajima2017PhysRevA.95.043625,horikoshi2019cold}.

In this paper, we examine the effective-range correction in the density-induced BEC-BCS crossover \red{in three dimensions} by using the Hartree-Fock-Bogoliubov (HFB) theory,
which takes both mean fields in the Cooper and density channels self-consistently~\cite{Urban2021PhysRevA.103.063306}, because the mean-field shift for the density is not negligible in the case with finite-range interactions in contrast to the contact-type one.
We employ the separable interaction based on the $s$-wave phase shift~\cite{Inotani2020PhysRevC.102.065802,Urban2021PhysRevA.103.063306}.
Because we are interested in the {\it pure} effective-range corrections without any higher-order coefficients,
we use the separable interaction exactly reproducing the phase shift with the effective-range expansion up to $O(k^2)$~\cite{Wyk2018PhysRevA.97.013601}.
We discuss how the effective-range corrections affect superfluid properties and the sound velocity in the density-induced BEC-BCS crossover.
In addition, as a related topic, we also examine the high-momentum behavior of the momentum distribution function, which is known to be related to Tan's contact~\cite{TAN20082952,TAN20082971,TAN20082987} in the case with the contact-type interaction.
The contact parameter is now interested in nuclear systems in terms of short-range correlations as called nuclear contact~\red{\cite{WEISS2018211,cruz2021many}}.
The relation of these quantities between nuclei and cold atoms has been discussed in Ref.~\cite{Hen2015PhysRevC.92.045205}.
We also discuss whether the high-momentum tail associated with Tan's contact can be visible in the presence of the nonzero effective range.

This paper is organized as follows.
In Sec.~\ref{sec:2},
we explain the HFB theory for two-component fermions with finite-range two-body interaction.
Also, the details of the separable interaction and the equation for physical quantities are presented.
In Sec.~\ref{sec:3},
we show the numerical results of the superfluid order parameter, sound velocity, and the high-momentum tail of the momentum distribution and discuss the effective range correction.
In Sec.~\ref{sec:4}, we summarize the contents.
In what follows, we take $\hbar=k_{\rm B}=1$.
\red{In this paper, we consider the thermodynamic limit with an infinitely large volume $\mathcal{V}$. While we calculate the internal energy density $E=\mathcal{E}/\mathcal{V}$ and the number density $\rho=N/\mathcal{V}$ (where $\mathcal{E}$ and $N$ are the internal energy and the particle number, respectively),
we omit $\mathcal{V}$ in the following for convenience because the results do not depend on $\mathcal{V}$ in the thermodynamic limit.}

\section{Formulation}
\label{sec:2}
\subsection{HFB theory}
We consider a two-component fermions with finite-range interaction \red{in three dimensions} as
$H=H_0+V$, where
\begin{align}
    H_0&=\sum_{\bm{k},\sigma}\xi_{\bm{k}}c_{\bm{k},\sigma}^\dag c_{\bm{k},\sigma}
\end{align}
is the kinetic term with the single-particle energy $\xi_{\bm{k}}=k^2/(2m)-\mu$ measured from the chemical potential $\mu$ ($m$ is a mass).
The interaction term reads
\begin{align}
\label{eq:V1}
    V&=\sum_{\bm{k},\bm{k}',\bm{P}}
    U(\bm{k},\bm{k}')
    c_{\bm{k}+\frac{\bm{P}}{2},\uparrow}^\dag
    c_{-\bm{k}+\frac{\bm{P}}{2},\downarrow}^\dag
    c_{-\bm{k}'+\frac{\bm{P}}{2},\downarrow}
    c_{\bm{k}'+\frac{\bm{P}}{2},\uparrow},
\end{align}
where the coupling strength $U(\bm{k},\bm{k}')$ depends on the relative momenta of two fermions.

%We decompose $V$ into the Cooper and density channels as
%\begin{align}
%\label{eq:V2}
%    V&\simeq\sum_{\bm{k},\bm{k}'}
%    U(\bm{k},\bm{k}')
%    c_{\bm{k},\uparrow}^\dag
%    c_{-\bm{k},\downarrow}^\dag
%    c_{-\bm{k}',\downarrow}
%    c_{\bm{k}',\uparrow} \cr%\quad (\bm{P}=\bm{0}, \bm{k}\neq\bm{k}')\cr
%    &+\sum_{\bm{p},\bm{p}'}
%    U\left(\frac{\bm{p}-\bm{p}'}{2},\frac{\bm{p}-\bm{p}'}{2}\right)
%    c_{\bm{p},\uparrow}^\dag
%    c_{\bm{p},\uparrow}
%    c_{\bm{p}',\downarrow}^\dag
%    c_{\bm{p}',\downarrow},
    %&+\sum_{\bm{k},\bm{P}}
    %U(\bm{k},\bm{k})
    %c_{\bm{k}+\bm{P}/2,\uparrow}^\dag
    %c_{\bm{k}+\bm{P}/2,\uparrow}
    %c_{-\bm{k}+\bm{P}/2,\downarrow}^\dag
    %c_{-\bm{k}+\bm{P}/2,\downarrow} \quad (\bm{P}\neq\bm{0},\bm{k}=\bm{k}').
%\end{align}
%where we assumed $\bm{P}\simeq\bm{0}$ and $\bm{k}\simeq\bm{k}'$ in the first and second terms of Eq.~(\ref{eq:V2}), and took $\bm{p}=\bm{k}+\bm{P}/2$ and $\bm{p}'=-\bm{k}+\bm{P}/2$ for the second term.
%We note that while the two terms in Eq.~(\ref{eq:V2}) originates from Eq.~(\ref{eq:V1}), there are no double countings because they belong to the different domains of the momentum spaces in Eq.~(\ref{eq:V1}) (i.e., $\bm{k}$, $\bm{k}'$, and $\bm{P}$).
We apply the mean-field approximation in each term by introducing the pairing gap
\begin{align}
\label{eq:3}
    \Delta(\bm{k})=-\sum_{\bm{k}'}U(\bm{k},\bm{k}')\langle c_{-\bm{k}',\downarrow} c_{\bm{k}',\uparrow} \rangle,
\end{align}
and the HF-like self-energy
\begin{align}
\label{eq:4}
    \Sigma_{\sigma}(\bm{p})
    =\sum_{\bm{p}'}
        U\left(\frac{\bm{p}-\bm{p}'}{2},\frac{\bm{p}-\bm{p}'}{2}\right)
    \langle
    c_{\bm{p}',\bar{\sigma}}^\dag
    c_{\bm{p}',\bar{\sigma}} \rangle,
\end{align}
where $\bar{\sigma}$ denotes the opposite spin of $\sigma$.
\red{In Eqs.~(\ref{eq:3}) and (\ref{eq:4}), $\langle\cdots\rangle$ denotes the expectation value with respect to the thermal equilibrium.}
Using these momentum-dependent mean fields, we obtain 
\begin{align}
    H_{\rm HFB}
    &=\sum_{\bm{k},\sigma}\left[\xi_{\bm{k}}+\Sigma_{\sigma}(\bm{k})\right]c_{\bm{k},\sigma}^\dag c_{\bm{k},\sigma}\cr
    &-\sum_{\bm{k}}\left[\Delta^*(\bm{k})c_{-\bm{k},\downarrow}c_{\bm{k},\uparrow}+\Delta(\bm{k})c_{\bm{k},\uparrow}^\dag c_{-\bm{k},\downarrow}^\dag\right]\cr
    &-\sum_{\bm{k},\bm{k}'}U(\bm{k},\bm{k}')
    \langle c_{\bm{k},\uparrow}^\dag c_{-\bm{k},\downarrow}^\dag \rangle
    \langle c_{-\bm{k}',\downarrow} c_{\bm{k}',\uparrow}\rangle\cr
    &-\sum_{\bm{p},\bm{p}'}
    U\left(\frac{\bm{p}-\bm{p}'}{2},\frac{\bm{p}-\bm{p}'}{2}\right)
    \langle c_{\bm{p},\uparrow}^\dag c_{\bm{p},\uparrow}\rangle
    \langle c_{\bm{p}',\downarrow}^\dag c_{\bm{p}',\downarrow}\rangle,
\end{align}
which is referred to as the mean-field Hamiltonian of the HFB theory in Ref.~\cite{Urban2021PhysRevA.103.063306}.

Hereafter, we consider the spin-balanced case ($\Sigma_{\uparrow}=\Sigma_{\downarrow}$)
Using the Nambu spinor $\Psi_{\bm{k}}=(c_{\bm{k},\uparrow} \quad c_{-\bm{k},\downarrow}^\dag)^{\rm T}$, one can rewrite $H_{\rm HFB}$ as
\begin{align}
    H_{\rm HFB}
    &=\sum_{\bm{k}}\Psi_{\bm{k}}^\dag
    \left[\xi_{\bm{k}}\tau_3+\Sigma(\bm{k})\tau_3-\Delta(\bm{k})\tau_1\right]
    \Psi_{\bm{k}}\cr
        &+\sum_{\bm{k}}\Delta(\bm{k})
    \langle c_{\bm{k},\uparrow}^\dag c_{-\bm{k},\downarrow}^\dag \rangle
    -\sum_{\bm{p}}
    \Sigma(\bm{p})
    \langle c_{\bm{p},\uparrow}^\dag c_{\bm{p},\uparrow}\rangle\cr
    &+\sum_{\bm{k}}[\xi_{\bm{k}}+\Sigma(\bm{k})],
\end{align}
where we omitted the spin indices in the HF-like self-energy $\Sigma(\bm{k})$.
\red{$\tau_{j=1,2,3}$ is the Pauli matrix acting on the Nambu spinor.}
Taking the Bogoiubov transformation, we obtain the ground-state energy \red{density} $E$ as
\begin{align}
    E&=\sum_{\bm{k}}\left[\xi_{\bm{k}}+\Sigma(\bm{k})-E_{\bm{k}}\right]\cr
    &+\sum_{\bm{k}}\Delta(\bm{k})
    \langle c_{\bm{k},\uparrow}^\dag c_{-\bm{k},\downarrow}^\dag \rangle
    -\sum_{\bm{p}}
    \Sigma(\bm{p})
    \langle c_{\bm{p},\uparrow}^\dag c_{\bm{p},\uparrow}\rangle,
\end{align}
where $E_{\bm{k}}=\sqrt{\{\xi_{\bm{k}}+\Sigma(\bm{k})\}^2+|\Delta(\bm{k})|^2}$ is the quasiparticle dispersion.

\subsection{Separable interaction with effective range}
In this paper, to consider effects of nonzero effective range,
we introduce the separable $s$-wave interaction
\begin{align}
    U(\bm{k},\bm{k}')=g\gamma_{k}\gamma_{k'}.
\end{align}
The two-body $T$-matrix in the center-of-mass frame is given by~\cite{GURARIE20072}
\begin{align}
    T(\bm{k},\bm{k}';\omega)&=U(\bm{k},\bm{k}')+\sum_{\bm{p}}
    U(\bm{k},\bm{p})\frac{1}{\omega_+-2\varepsilon_{\bm{p}}}
    T(\bm{p},\bm{k}';\omega),
\end{align}
where $\omega_+=\omega+i0$.
Assuming the separable form of the $T$-matrix, we obtain
\begin{align}
\label{eq:tmatrix}
    T(\bm{k},\bm{k}';\omega)=\gamma_k\left[\frac{1}{g}-\sum_{\bm{p}}
    \frac{\gamma_p^2}{\omega_+-p^2/m}\right]^{-1}\gamma_{k'}.
\end{align}
The onshell $T$-matrix is associated with the $s$-wave phase shift $\delta_s(k)$ as
\begin{align}
\label{eq:pshift}
    -\frac{m}{4\pi}T(\bm{k},\bm{k};2\varepsilon_{\bm{k}})=\frac{1}{k\cot\delta_s(k)-ik}.
\end{align}
$k\cot\delta_s(k)$ can be expanded with respect to $k$ as
\begin{align}
\label{eq:effran}
    \red{k\cot}\delta_s(k)=-\frac{1}{a}+\frac{1}{2}rk^2 -\mathcal{S}r^3k^4 +O(k^5),
\end{align}
where $a$, $r$, and $\mathcal{S}$ are the scattering length, effective range, and shape parameter~\cite{Hamada1961PTP.26.153}, respectively.
Using Eqs.~(\ref{eq:tmatrix}), (\ref{eq:pshift}), and (\ref{eq:effran}),
we obtain the relation between $U(\bm{k},\bm{k}')$ and the low energy constants given by~\cite{Inotani2020PhysRevC.102.065802,Urban2021PhysRevA.103.063306}
\begin{align}
\label{eq:a}
    \frac{m}{4\pi a}=\frac{1}{\gamma_{k=0}^2}
    \left[\frac{1}{g}+\frac{m}{2\pi^2}\int_{0}^{\infty}dk\gamma_k^2\right],
\end{align}
\begin{align}
\label{eq:r}
    &-\frac{m}{4\pi}\left(\frac{rk^2}{2}-\mathcal{S}r^3k^4\right)+O(k^5)=\frac{1}{g}\left[\frac{1}{\gamma_k^2}-\frac{1}{\gamma_{k=0}^2}\right]\cr
    &\quad\quad\quad\quad\quad
    \red{+\frac{m}{2\pi^2}\int_{0}^{\infty}dp
    \left[\frac{p^2\gamma_{p}^2-k^2\gamma_{k}^2}{p^2-k^2}
    -\frac{\gamma_p^2}{\gamma_{k=0}^2}
    \right]},
\end{align}
Because we are interested in the effective-range corrections,
$\mathcal{S}$ is chosen as zero.
In such a case, the fact that the leading term in the right hand side of Eq.~(\ref{eq:r}) is proportional to $k^2$ motivates us to use the form factor given by~\cite{Nozieres1985bose,Wyk2018PhysRevA.97.013601,Tajima2019JPSJ}
\begin{align}
\label{eq:gamma}
    \gamma_{k}=\frac{1}{\sqrt{1+(k/\Lambda)^2}},
\end{align}
where $\Lambda$ corresponds to the high-momentum cutoff.
Note that if one uses the form factor given by $\gamma_{k,{\rm Y}}=[1+(k/\Lambda_{\rm Y})^2]^{-1}$ (referred to as Yamaguchi potential~\cite{Yamaguchi1954PhysRev.95.1628}),
one obtains nonzero $\mathcal{S}$, in contrast with the present case (see Appendix~\ref{app:A}).
Eventually, the parameters in $U(\bm{k},\bm{k}')$ are expressed by
\begin{align}
\label{eq:cutoff}
    \Lambda=\frac{1}{r}\left[1+\sqrt{1-\frac{2r}{a}}\right],
\end{align}
\begin{align}
\label{eq:coupling}
    g=\frac{4\pi a}{m}\frac{1}{1-a\Lambda}.
\end{align}
We note that, if $r/a$ approaches $0.5$, $\Lambda$ becomes complex where a physical bound state merges with a spurious state~\cite{ebert2021alternative}.
In this paper, we discuss the region $r/a<0.5$ avoiding such a singularity.
In Appendix~\ref{app:A}, we summarize other separable potentials such as gaussian form factor (which is also used as a regulator in a chiral effective field theory~\cite{MACHLEIDT20111}).
In Appendix~\ref{app:B}, we discuss the relation with the screened Coulomb interaction which is relevant to an electron-hole system.
While the beyond-mean-field effect such as pairing and density fluctuations has been discussed in Ref.~\cite{Wyk2018PhysRevA.97.013601,Urban2021PhysRevA.103.063306,Inotani2020PhysRevC.102.065802},
the mean-field framework is sufficient to see a qualitative behavior in the BEC-BCS crossover regime with finite effective range~\cite{Andrenacci1999PhysRevB.60.12410,Parish2005PhysRevB.71.064513}.

\subsection{Equations for the numerical calculation of physical quantities}
In the case with separable interactions,
the pairing gap can be rewritten as
\begin{align}
    \Delta(\bm{k})=-\gamma_k g\sum_{\bm{k}'}\gamma_{k'}
    \langle c_{-\bm{k}',\downarrow} c_{\bm{k}',\uparrow} \rangle
    \equiv \Delta\gamma_{k}.
\end{align}
The magnitude of the superfluid order parameter $|\Delta|$ is determined via the saddle-point condition $\frac{\partial E}{\partial |\Delta|^2}=0$ (gap equation) where
\begin{align}
    E&=\sum_{\bm{k}}\left[\xi_{\bm{k}}+\Sigma(\bm{k})-E_{\bm{k}}\right]-\frac{|\Delta|^2}{g}
    -\sum_{\bm{k}}
    \Sigma(\bm{k})
    n_{\bm{k}}.
\end{align}
The explicit form of the gap equation reads
\begin{align}
\label{eq:gapeq}
    0&=\frac{1}{g}+\sum_{\bm{k}}\frac{\gamma_{k}^2}{2E_{\bm{k}}}\cr
    &\equiv\frac{m}{4\pi a}+\sum_{\bm{k}}\gamma_k^2
    \left[\frac{1}{2E_{\bm{k}}}-\frac{m}{k^2}\right].
\end{align}
Also, $\Sigma(\bm{k})$ is in the form of
\begin{align}
\label{eq:sig}
    \Sigma(\bm{k})
    =g\sum_{\bm{k}'}\gamma_{\frac{|\bm{k}-\bm{k}'|}{2}}^2
    n_{\bm{k}'},
\end{align}
where 
\begin{align}
\label{eq:momd}
    n_{\bm{k}}=
    \frac{1}{2}\left[1-\frac{\xi_{\bm{k}}+\Sigma(\bm{k})}{E_{\bm{k}}}\right]
    ,
\end{align}
is the momentum distribution.

To obtain the density dependence of physical quantities,
we also need to solve number density equation
\begin{align}
\label{eq:numeq}
    \rho&
    =\sum_{\bm{k}}
    \left[1-\frac{\xi_{\bm{k}}+\Sigma(\bm{k})}{E_{\bm{k}}}\right],
\end{align}
with respect to $\mu$.
Furthermore, to see the mechanical stability of the system, we calculate the compressibility $\kappa$ as
\begin{align}
    \kappa=\frac{1}{\rho^2}\left(\frac{\partial \rho}{\partial \mu}\right).
\end{align}
Indeed, $\kappa$ is associated with the second derivative of the Helmholtz free energy $\frac{\partial^2 F}{\partial \rho^2}$ investigated in Ref.~\cite{Schonenberg2017PhysRevA.95.013633}.
When the system reaches the mechanical collapse, we obtain $\frac{\partial^2 F}{\partial \rho^2}=0$, namely, the divergent compressibility~\cite{seo2011compressibility}.
Moreover, it can be expressed in terms of the sound velocity $c_s$ as
\begin{align}
    c_s=\sqrt{\frac{1}{m\rho\kappa}}.
\end{align}
Hence, $c_s$ becomes zero when the mechanical collapse occurs.

In this way, we solve the gap equation (\ref{eq:gapeq}) and the number density equation (\ref{eq:numeq}) with monitoring $c_s$ or $\kappa$.
In addition, because the left hand side of Eq.~(\ref{eq:sig}) involves $\Sigma(\bm{k})$ in $n_{\bm{k}}$ as we showed in Eq.~(\ref{eq:momd}),
we need to self-consistently solve Eq.~(\ref{eq:sig}) with respect to $\Sigma(\bm{k})$.
In this study, we work on the case with positive $r$.
We note that to extend our study to the negative $r$, it is necessary to generalize the Hamiltonian to the coupled channel model~\cite{GURARIE20072,OHASHI2020103739,PhysRevA.97.043613,Tajima2019JPSJ,PhysRevA.101.013615}, which is out of scope in this paper.

\section{Results}
\label{sec:3}
\subsection{Superfluid order parameter}
\begin{figure}[t]
    \centering
    \includegraphics[width=8cm]{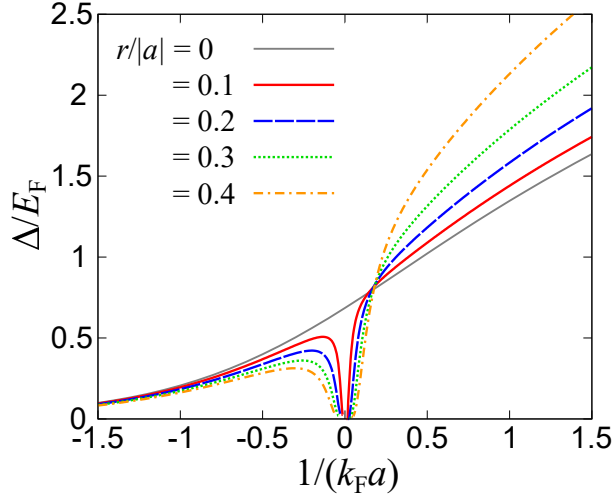}
    \caption{Superfluid order parameter $\Delta$ along the BCS-BEC crossover with nonzero effective range~$r$.}
    \label{fig:2}
\end{figure}
Figure~\ref{fig:2} shows the superfluid order parameter $\Delta/E_{\rm F}$ as a function of inverse scattering length $1/(k_{\rm F}a)$ with several ratios between $r$ and $a$, where $k_{\rm F}=\left(3\pi^2\rho\right)^{\frac{1}{3}}$ and $E_{\rm F}=\frac{k_{\rm F}^2}{2m}$ are the Fermi momentum and Fermi energy, respectively.
Here, $\Delta$ is taken as a real value without loss of generality.
The zero-range result ($r/a=0$) exhibits the continuous crossover from the BCS-type pairing in the weak-coupling regime [$1/(k_{\rm F}a)\lesssim -1$]
to the molecular BEC state in the strong-coupling regime [$1/(k_{\rm F}a)\gesim 1$].
Such a situation is indeed realized in an ultracold Ferm gases near the Feshbach resonance, where only $a$ is dramatically enhanced~\cite{Chin2010RevModPhys.82.1225} by tuning the external magnetic field.

On the other hand, in the case of the density-induced BEC-BCS crossover,
$k_{\rm F}$ can change instead of $a$.
When $a>0$ ($a<0$), the dimensionless coupling parameter $1/(k_{\rm F}a)$ runs from the strong-coupling regime, $1/(k_{\rm F}a)=\infty$ (the weak-coupling regime, $1/(k_{\rm F}a)=-\infty$), to the unitary limit ($1/(k_{\rm F}a)=0$), with the evolution of $k_{\rm F}\propto \rho^{\frac{1}{3}}$ as shown in Fig.~\ref{fig:1}.
However, in the unitary limit with the density evolution, the effective range correction cannot be negligible even if $r$ is much less than $a$, as characterized by the magnitude of the dimensionless parameter $k_{\rm F}r$.
Namely, $k_{\rm F}r$ goes infinity with $1/(k_{\rm F}a)\rightarrow 0$ in the high-density limit.
In this regard, the finite-range effects should be seriously examined in the high-density regime compared to the value of the inverse scattering length $1/(k_{\rm F}a)\simeq 0$.
As one can see in Fig.~\ref{fig:2}, $\Delta$ is strongly suppressed in the region with $1/(k_{\rm F}a)\simeq 0$, that is, $k_{\rm F}r\rightarrow\infty$, due to the effective range correction, regardless of the sign of $a$.
Such a behavior is consistent with the generalized crossover scenario~\cite{Tajima2019JPSJ}, where the crossover boundary is qualitatively given by $\delta_s(k=k_{\rm F})=0$ instead of $1/(k_{\rm F}a)=0$ for the zero-range interaction.

\red{In contrast to the region near $1/(k_{\rm F}a)=0$, one can see that $\Delta$ is strongly enhanced by the finite-range correction in the diulte BEC regime $1/(k_{\rm F}a)\gesim 0.5$. This is associated with the increase of the two-body binding energy $E_{\rm b}$ as~\cite{Tajima2019JPSJ}
\begin{align}
    E_{\rm b}&=\frac{1}{ma^2}\frac{1}{[1-1/(a\Lambda)]^2}\cr
    &\equiv\frac{1}{mr^2}\left(1-\sqrt{1-\frac{2r}{a}}\right)^2.
\end{align}
Indeed, $E_{\rm b}$ reaches $\frac{4}{ma^2}$ at $r=a/2$,
which is four times the zero-range counterpart $\frac{1}{ma^2}$.
In such a strong-coupling regime with $\mu<0$, we obtain the approximate form of $\Delta$ as (see more details in Appendix~\ref{app:C})
\begin{align}
\label{eq:gapBEC}
    \Delta
    &\simeq\sqrt{\frac{16}{3\pi}}E_{\rm F}^{3/4}\left(\frac{E_{\rm b}}{2}\right)^{1/4}
    \left(1+\frac{\sqrt{mE_{\rm b}}}{\Lambda}\right).
\end{align}
In this way,
the increase of $E_{\rm b}$ induced by the effective-range correction leads to the enhancement of $\Delta$ in the dilute BEC regime. 
}

\begin{figure}[t]
    \centering
    \includegraphics[width=8cm]{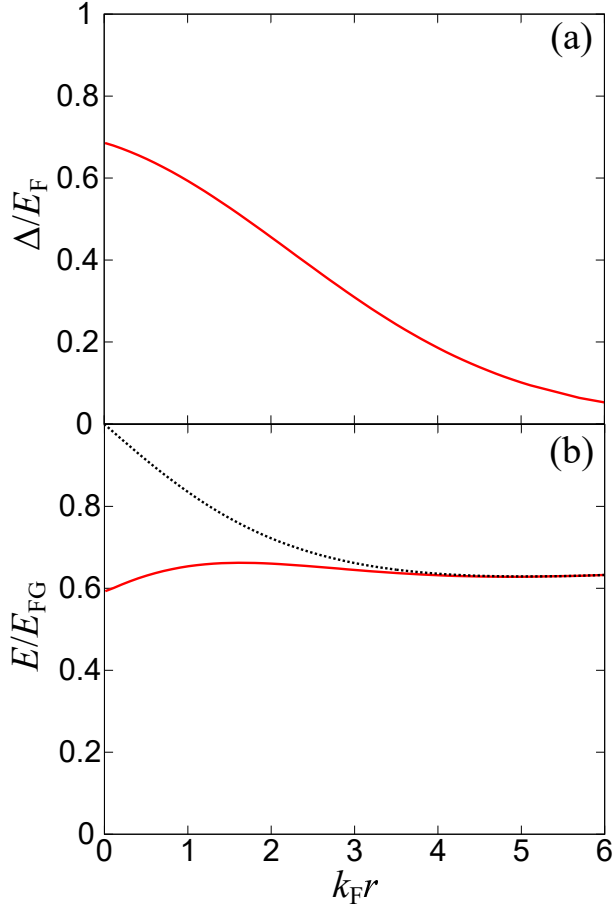}
    \caption{(a)Superfluid order parameter and (b) \red{internal} energy \red{density} as a function of the dimensionless range parameter $k_{\rm F}r$ at $1/(k_{\rm F}a)=0$. The dotted line shows the result with $\Delta=0$ in the normal phase.}
    \label{fig:3}
\end{figure}

To see the role of the effective range in the high-density regime,
we plot $\Delta/E_{\rm F}$ and $E/E_{\rm FG}$ as functions of $k_{\rm F}r$ at $1/(k_{\rm F}a)=0$ in Fig.~\ref{fig:3}, where $E_{\rm FG}=\frac{3}{5}\rho E_{\rm F}$ is the non-interacting counterpart of the \red{internal} energy \red{density at $T=0$}.
\red{In Fig.~\ref{fig:3}, the results depend on only $k_{\rm F}r$ because of $|a|=\infty$. Hence, the horizontal axis of Fig.~\ref{fig:3} can be regarded as either $k_{\rm F}$-dependence with fixed $r$ or $r$-dependence with fixed $k_{\rm F}$. This parameter choice is useful to examine $k_{\rm F}r$-dependence of the results in the crossover regime.}
One can see that $\Delta$ decreases monotonically with increasing $k_{\rm F}r$ because the positive effective range induces the momentum cutoff $\Lambda$ in the Cooper channel $U(\bm{k},\bm{k})$.
Such a behavior is consistent with the previous work~\cite{Wyk2018PhysRevA.97.013601,Schonenberg2017PhysRevA.95.013633,Tajima2019JPSJ}.
\red{While the decrease of $\Delta$ leads to the increase of $E$ (corresponding to the suppression of the condensation energy with increasing $k_{\rm F}r$), the suppression of $E$ by the HF self-energy becomes strong with increasing $k_{\rm F}r$.
The competition between these two effects leads to the peak structure of $E$ in Fig.~\ref{fig:3}(b).}

However, we do not find any phase transitions with increasing $k_{\rm F}r$ in our model with the {\it pure} effective range correction, in contrast to Ref.~\cite{Schonenberg2017PhysRevA.95.013633} as we will discuss later.
$E$ is always less than the result without the pairing gap.
The results with and without the pairing gap become close to each other at larger $k_{\rm F}r$.
Although we do not explicitly show the limit of $k_{\rm F}r\rightarrow \infty$, one can analytically find that the ideal gas result can be realized in this limit for our model. 
At $1/(k_{\rm F}a)=0$, we obtain $\Lambda=\frac{2}{r}$ and $g=-\frac{4\pi}{m\Lambda}$ from Eqs.~(\ref{eq:cutoff}) and (\ref{eq:coupling}).
In this regard, we obtain
\begin{align}
\label{eq:Urlarge}
    U(\bm{k},\bm{k})\rightarrow-\frac{4\pi^2}{m}\delta(k) \quad (r\rightarrow \infty),
\end{align}
where the delta function $\delta(k)$ for the relative momenta in Eq.~(\ref{eq:Urlarge}) indicates the infinitely long-range constant interaction in the coordinate space when performing the Fourier transformation.
Indeed, Eq.~(\ref{eq:Urlarge}) can also be obtained by using the gaussian form factor $\gamma_{k,{\rm G}}=e^{-k^2/\Lambda_{\rm G}^2}$ in the limit of $r\rightarrow\infty$ with finite $a<0$ (see Appendix~\red{\ref{app:A}}).
Because $\Delta$ disappears in the limit of $r\rightarrow \infty$,
the HF self-energy $\Sigma_{r\rightarrow\infty}(\bm{k})$ is given by
\begin{align}
    \Sigma_{r\rightarrow\infty}(\bm{k})&=g\sum_{\bm{k}'}\gamma_{\frac{|\bm{k}-\bm{k}'|}{2}}^2
    \theta\left(-\frac{k'^2}{2m}+\mu-\Sigma(\bm{k}')\right),
\end{align}
where $\theta(x)$ is the Heaviside step function associated with the Fermi distribution function.
One can analytically obtain the vanishing shift as
\begin{align}
\label{eq:sig0}
    \Sigma_{r\rightarrow\infty}(\bm{k})
    &=-\frac{2}{mk}\int_{0}^{\infty}k'dk'\int_{|k-k'|}^{|k+k'|}ds
    \delta(s)s\cr
    &\times\theta\left(-\frac{k'^2}{2m}+\mu-\Sigma(\bm{k}')\right)\cr
    &=0,
\end{align}
where we took $s=\sqrt{k^2+k'^2-2kk'\cos\theta_{\bm{k}\bm{k}'}}$ ($\theta_{\bm{k}\bm{k}'}$ is the angle between $\bm{k}$ and $\bm{k}'$) and used $s\delta(s)=0$.
In this way, one can expect that the interaction effect is negligible in the limit of $k_{\rm F}r\rightarrow \infty$. 

\begin{figure}
    \centering
    \includegraphics[width=8cm]{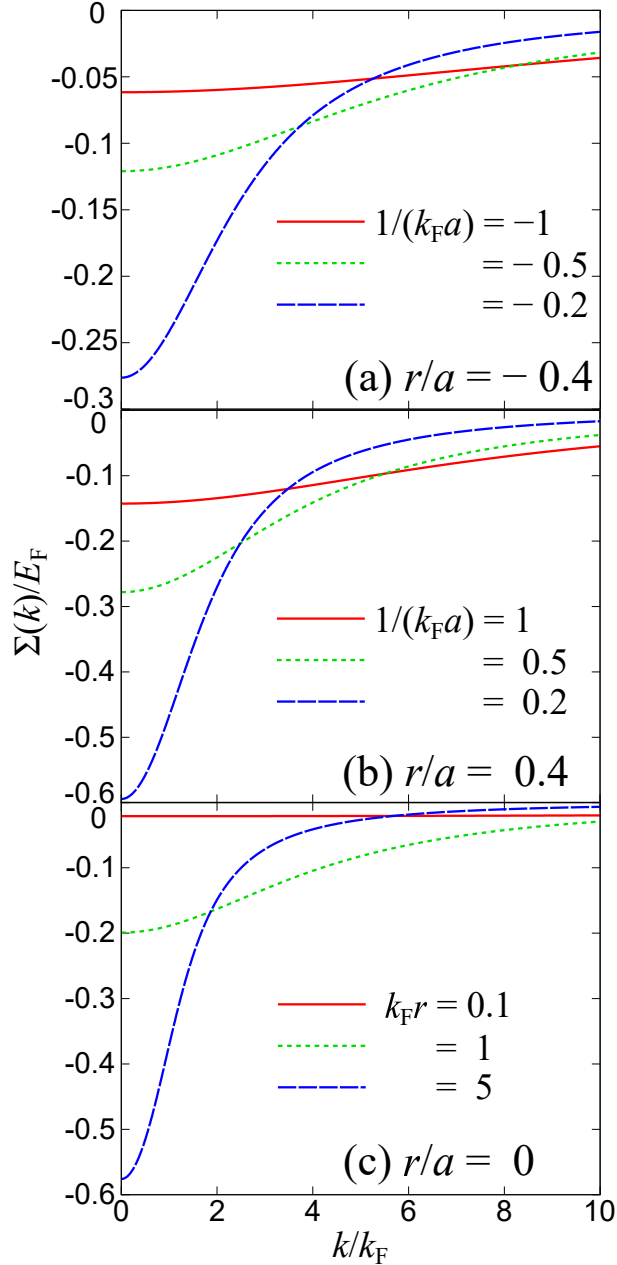}
    \caption{HF self-energies $\Sigma(k)$ at (a) $r/a=-0.4$ ($a<0$), (b) $r/a=0.4$ ($a>0$), and (c) $r/a=0$ ($|a|=\infty$).}
    \label{fig:4}
\end{figure}
Figure~\ref{fig:4} shows calculated $\Sigma(k)$ at (a) $r/a=-0.4$ ($a<0$), (b) $r/a=0.4$ ($a>0$), and (c) $r/a=0$ ($|a|=\infty$) (note that $\Sigma$ depends only on $k=|\bm{k}|$ because of the spherical symmetry).
\red{We note that in Fig.~\ref{fig:4}(c) $|a|=\infty$ is fixed to examine $k_{\rm F}r$-dependence of $\Sigma(k)$.}
In the panels (a) and (b) of Fig.~\ref{fig:4}, the dimensionless parameter $1/(k_{\rm F}a)$ is varied with fixed $r/a$.
Thus, $k_{\rm F}r=\frac{r/a}{(k_{\rm F}a)^{-1}}$ becomes large when $1/(k_{\rm F}a)$ approaches zero (see also Fig.~\ref{fig:1}).
Generally, the magnitude of $\Sigma(k)$ becomes large in the low-momentum regime ($k\simeq 0$)
and moreover enhanced with increasing $k_{\rm F}r$,
reflecting the structure of the interaction.
Indeed, this enhancement is associated with the normalized coupling $g\frac{mk_{\rm F}}{2\pi^2}\simeq -\frac{k_{\rm F}r}{\pi}$.
The momentum dependence of $\Sigma(k)$ is also affected by $k_{\rm F}r$ as $\Sigma(k)$ is suppressed in the high-momentum regime ($k/k_{\rm F}\gesim \Lambda/k_{\rm F}\simeq \frac{2}{k_{\rm F}r}$). The panel (c) of Fig.~\ref{fig:4} shows the explicit $k_{\rm F}r$ dependence of $\Sigma(k)$ at $r/a=0$.
One can see that $\Sigma(k)$ clearly shrinks in the momentum space. 
In the high-density regime, the momenta near the Fermi surface (i.e., $k\simeq k_{\rm F}$) is relevant for many-body physics and the \red{enhanced} behavior at $k=0$ is less important.
This result can also be understood from the fact that the contribution \red{near} $\bm{k}=\bm{0}$ is not visible in the momentum distribution $n_{\bm{k}}=\theta\left(-\frac{k^2}{2m}+\mu-\Sigma(\bm{k})\right)$ 
in the normal phase.

\begin{figure}
    \centering
    \includegraphics[width=8cm]{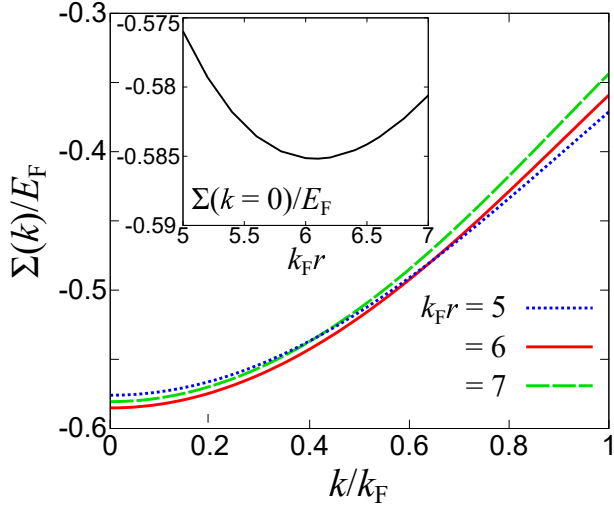}
    \caption{\red{HF self-energies $\Sigma(k)$ at $k_{\rm F}r=5$, $6$, and $7$ (where $|a|=\infty$).
    The inset shows the zero-momentum limit $\Sigma(k=0)$ as a function of $k_{\rm F}r$.}
    }
    \label{fig:5}
\end{figure}
\red{To see how the HF self-energy vanishes in the large effective-range regime as found in Eq.~(\ref{eq:sig0}), we plot $\Sigma(k)$ with larger $k_{\rm F}r$ in Fig.~\ref{fig:5}.
One can see that the absolute value of $\Sigma(k)$ at $k\simeq k_{\rm F}$ is suppressed at larger $k_{\rm F}r$, indicating shrinking of $\Sigma(k)$ in the momentum space.
Moreover, the absolute value of the zero-momentum shift $\Sigma(k=0)$, which gives the largest magnitude for a given $k_{\rm F}r$ in the momentum space,
also starts to decrease around $k_{\rm F}r=6.1$, 
as shown in the inset of Fig.~\ref{fig:5}.
}
Therefore, the results at $k_{\rm F}r\rightarrow\infty$ can approach the ideal gas results except for the presence of other corrections such as $\mathcal{S}$, higher partial waves, short-range repulsion, and three-body forces.  
It is in contrast to Ref.~\cite{Schonenberg2017PhysRevA.95.013633} where the mechanical collapse was reported in the large effective-range region.
While we are not in the position to clarify the differences between our study and Ref.~\cite{Schonenberg2017PhysRevA.95.013633},
the large effective range limit may be sensitive to the small deviation of the phase shift in the high-momentum region and generally the higher partial waves such as $p$- and $d$-waves cannot be negligible there.
\red{Because the pairing gap at large $k_{\rm F}r$ is so small that the HFB result is close to the HF result as shown in Fig.~\ref{fig:3},
the difference between our result and the previous work using both the HF and diffusion Monte Carlo method is not due to the many-body calculation but the interaction model.}
In this regard, our work examines purely effective range corrections without any other partial waves and short-range parameters.
Such a situation can be tested in the future cold atomic experiments by enhancing the effective range with the optical field method~\cite{Wu2012PhysRevA.86.063625,Wu2012PhysRevLett.108.010401,Arunkumar2018PhysRevLett.121.163404,Arunkumar2019PhysRevLett.122.040405}.

\subsection{Sound velocity}
\begin{figure}
    \centering
    \includegraphics[width=8cm]{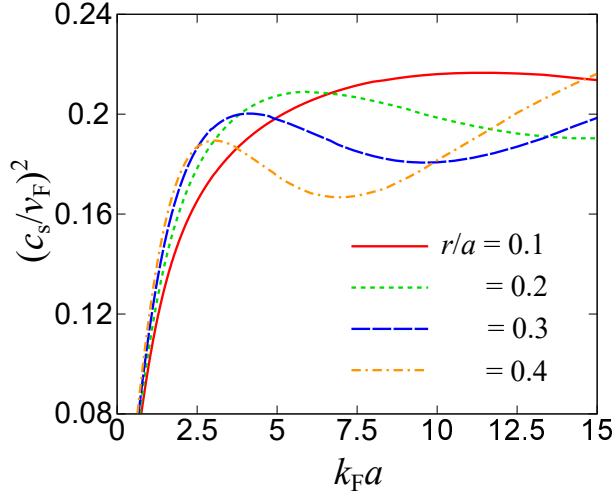}
    \caption{Squared sound velocity $c_s^2$ at positive scattering length ($a>0$). $v_{\rm F}=k_{\rm F}/m$ is the Fermi velocity.}
    \label{fig:6}
\end{figure}
\begin{figure}
    \centering
    \includegraphics[width=8cm]{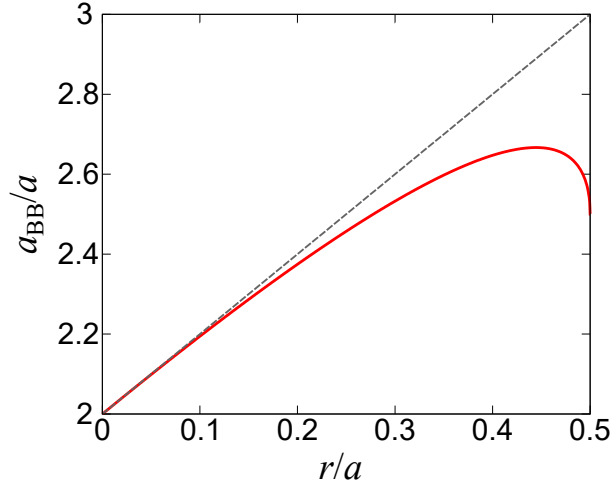}
    \caption{\red{Effective-range dependence of the molecule-molecule scattering length $a_{\rm BB}$ within the HFB theory in the BEC limit. The dashed line shows the asymptotic behavior in the zero-range limit given by $a_{\rm BB}\simeq 2a+2r$.}}
    \label{fig:7}
\end{figure}

To see a mechanical stability of the system, 
we examine the sound velocity $c_s$, which is associated with the inverse compressibility.
Because we are interested in the density-induced BEC-BCS crossover \red{with fixed $a$ and $r$}, we focus on the case with the positive scattering length $a>0$ in the following.
 \red{We note that the BCS-BCS crossover can be realized at $a<0$, where the dilute BCS phase with $1/(k_{\rm F}a)\lesssim -1$ and $k_{\rm F}r\simeq 0$ changes into the dense BCS phase with $1/(k_{\rm F}a)\simeq 0$ and $k_{\rm F}r\gesim 1$ in the evolution of $k_{\rm F}$ along the line of $r/a<0$ in Fig.~\ref{fig:1}}.
Figure~\ref{fig:5} shows the density dependence of $c_s^2$ at the positive scattering length $a>0$, where $v_{\rm F}=k_{\rm F}/m$ is the Fermi velocity.
Thus, the low-density regime ($k_{\rm F}a\simeq 0$) is dominated by the two-body bound state formation.
Such a situation is similar to the density-induced BEC-BCS crossover discussed in condensed-matter~\cite{nakagawa2021gate,Suzuki2022PhysRevX.12.011016} and in nuclear systems~\cite{Kojo2022PhysRevD.105.076001}.
It is known that $c_s$ is given by the sound velocity of the Bogoliubov phonon $v_{\rm B}=\sqrt{\frac{U_{\rm BB}\rho_{\rm B}}{m_{\rm B}}}$~\cite{OHASHI2020103739},
where $U_{\rm BB}$, $\rho_{\rm B}=\rho/2$, and $m_{\rm B}=2m$ are the molecule-molecule repulsive interaction strength, the molecular density, and the molecular mass, respectively.
In this sense, $c_s$ increases with increasing $k_{\rm F}$ in the low-density regime \red{because of the molecule-molecule repulsion.
In the BEC limit, the gap equation given by Eq.~(\ref{eq:gapeq}) is equivalent to the Hugenholtz-Pines (HP) condition of molecular bosons within the mean-field level~\cite{PhysRevLett.91.030401,PhysRevB.98.104507}.
In this regard, one can extract the molecule-molecule scattering length $a_{\rm BB}$ from the HFB theory as (see also Appendix~\ref{app:C})
\begin{align}
\label{eq:abb}
    a_{\rm BB}=2a\frac{1+\frac{4\sqrt{mE_{\rm b}}}{\Lambda}}{\left(1+\frac{\sqrt{mE_{\rm b}}}{\Lambda}\right)^2},
\end{align}
where we used $U_{\rm BB}=\frac{4\pi a_{\rm BB}}{m_{\rm B}}$.
We note that Eq.~(\ref{eq:abb}) obtained in the HFB theory is not exact but consistent with the Born approximation with respect to $U_{\rm BB}$~\cite{OHASHI2020103739}.
Indeed, Eq.~(\ref{eq:abb}) reproduces $a_{\rm BB}=2a$ in the zero-range limit~\cite{PhysRevB.61.15370,doi:10.1143/JPSJ.74.2659}, whereas the exact value is given by $a_{\rm BB}=0.6a$~\cite{PhysRevLett.93.090404}.
However, it is useful to qualitatively examine the finite-range correction on the molecule-molecule repulsion.
Using $\frac{\sqrt{mE_{\rm b}}}{\Lambda}=\frac{1-\sqrt{1-2\frac{r}{a}}}{1+\sqrt{1-2\frac{r}{a}}}$,
we plot $a_{\rm BB}$ as a function of $r/a$ in Fig.~\ref{fig:7}.}
\red{It is found that the molecule-molecule repulsion becomes stronger with increasing $r/a$.
This result is consistent with the enhanced sound velocity with larger $r/a$ in the dilute BEC regime ($k_{\rm F}a\lesssim 2.5$).
In the small effective-range regime, $a_{\rm BB}$ linearly increases with $r$ as $a_{\rm BB}\simeq 2a+2r$.
$a_{\rm BB}$ exhibits a maximum value ($a_{\rm BB}\simeq2.67a$) at $r/a\simeq0.44$ and eventually approaches $2.5$ at $r/a=0.5$.
}

When the density is close to the region where the effective-range correction is significant (i.e., $k_{\rm F}r\equiv k_{\rm F}a\times\frac{r}{a}\simeq 1$),
$(c_s/v_{\rm F})^2$ exhibits a maximum and tends to decrease with $k_{\rm F}$.
This decreasing behavior is reminiscent of the precursor of mechanical collapse associated with the enhanced compressibility in terms of fermionic degrees of freedom, where $\mu$ becomes positive around $k_{\rm F}a=0.5$~\cite{BCS-BEC,OHASHI2020103739,STRINATI20181} (also see that pairing is strongly suppressed there as shown in Fig.~\ref{fig:2}).
Indeed, it can be qualitatively understood as the Stoner enhancement of the compressibility through the zero-momentum Hartree shift~\cite{Wyk2018PhysRevA.97.013601}
\begin{align}
\label{eq:Hartree}
    \Sigma_{\rm H}=U(\bm{0},\bm{0})\rho/2\equiv g\rho/2.
\end{align}
While $\Sigma_{\rm H}$ is valid in the lower-density region compared to $\Sigma(\bm{k})$ in Eq.~(\ref{eq:sig}),
Eq.~(\ref{eq:Hartree}) is nothing more than the usual form of the Hartree shift for long-range interactions without the partial-wave decomposition.%, such as Coulomb force~\cite{fetter2012quantum}.
In this case, the squared sound velocity $c_{s,{\rm H}}^2$ reads
\begin{align}
\label{eq:cs2h}
    c_{s,{\rm H}}^2= \frac{1}{m\rho\kappa_0}\left[1-\frac{2}{\pi}\frac{1}{\Lambda/k_{\rm F}-(k_{\rm F}a)^{-1}}\right],
\end{align}
where $\kappa_0=\frac{3}{2\rho E_{\rm F}}$ is the compressibility of an ideal Fermi gas, and $\Delta$ is taken to be zero for simplicity.
Moreover, in the high-density regime where $1/(k_{\rm F}a)\simeq 0$, one can find $c_{s,{\rm H}}^2\simeq \frac{1}{m\rho\kappa_0}\left(1-\frac{k_{\rm F}r}{\pi}\right)$,
which decreases with increasing $k_{\rm F}r$ and becomes zero at $k_{\rm F}r=\pi$.
To some extent this is the same order with the result in Ref.~\cite{Schonenberg2017PhysRevA.95.013633} where $c_{s}^2=0$ is found at $k_{\rm F}r=1.91$.

\begin{figure}
    \centering
    \includegraphics[width=8cm]{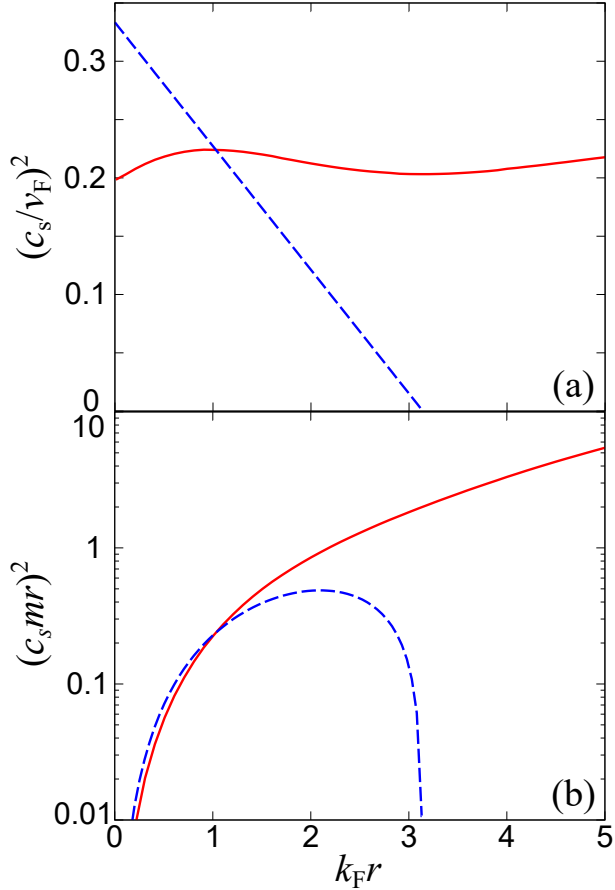}
    \caption{Square sound velocity $c_{s}^2$ as a function of the dimensionless range parameter $k_{\rm F}r$ at divergent scattering length.
    The panels (a) and (b) show $(c_s/v_{\rm F})^2$ and $(c_smr)^2$, respectively.
    The solid and dashed curves represent the result of the HFB theory and the zero-momentum Hartree approximation given by Eq.~(\ref{eq:cs2h}), respectively.
    }
    \label{fig:8}
\end{figure}

However, as we found in Eq.~(\ref{eq:sig0}),
the interaction effect becomes less relevant in the large effective-range regime.
Hence, as shown in Fig.~\ref{fig:5}, $c_s^2$ tends to increase again towards the ideal-gas result [$(c_s/v_{\rm F})^2=1/3$], at $k_{\rm F}r\equiv k_{\rm F}a\times\frac{r}{a}\simeq \pi$ where the zero-momentum Hartree approximation breaks down.
Figure~\ref{fig:8} shows the $k_{\rm F}r$ dependence of $c_s^2$ in different units [i.e., (a) $(c_s/v_{\rm F})^2$ and (b) $(c_smr)^2$].
For comparison, we plot the results of the HFB theory (solid curve) and 
Eq.~(\ref{eq:cs2h}) (dashed curve).
Indeed, one can see that $(c_s/v_{\rm F})^2$ has a peak around $k_{\rm F}r=1$ and a minimum around $k_{\rm F}r=\pi\simeq3.14$.

If we measure $c_s^2$ with the density-independent scale [i.e., $\red{1/(mr)^2}$],
the HFB result monotonically increases with $\rho$.
On the other hand,
the zero-momentum Hartree approximation given by Eq.~(\ref{eq:cs2h}) still shows a peak structure in the intermediate regime.
Although this can be an artefact of the approximation,
the same behavior has been reported in Ref.~\cite{Schonenberg2017PhysRevA.95.013633}.
In this sense, while the {\it pure} effective-range correction indeed softens the equation of state (and hence suppress the sound velocity) but does not induce the peak of the sound velocity within the density evolution as expected in hadron-quark crossover in two-color QCD~\cite{Kojo2022PhysRevD.105.076001},
the other non-universal short-range corrections naturally involved in realistic interaction potentials may cause such a sound-velocity peak.
However, because the sound velocity approaches the conformal limit (i.e., non-interacting limit, corresponding to $v_{\rm F}/\sqrt{3}$ in our model) with increasing the density in the two-color QCD~\cite{kojo2021qcd}, 
the non-monotonic behavior of $(c_s/v_{\rm F})^2$ in Fig.~\ref{fig:5},
(approaching the non-interacting counter part $(c_s/v_{\rm F})^2=1/3$ in the high-density limit), might give an important insight to understand the crossover phenomena. 
%More detailed investigations for the sound-velocity peak will be left for future work.

\subsection{High-momentum tail}
In the end of this section,
we discuss the high-momentum tail in the momentum distribution $n_{\bm{k}}$ in the presence of nonzero effective range.
\begin{figure}
    \centering
    \includegraphics[width=8.5cm]{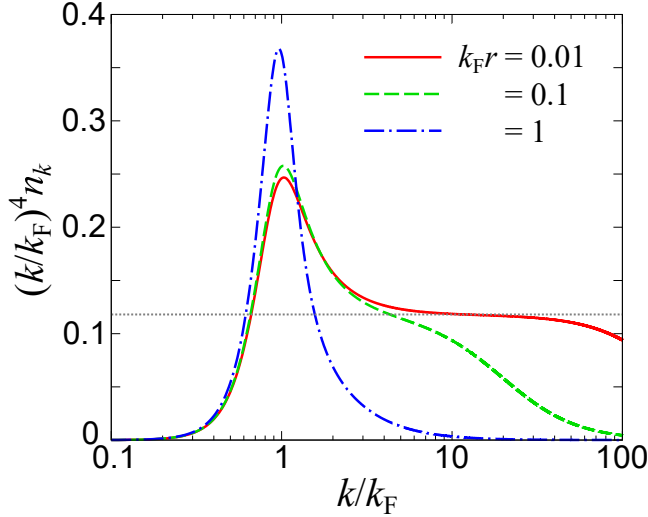}
    \caption{The high-momentum tail of the momentum distribution $n_k(k/k_{\rm F})^4$ at $a^{-1}=0$. The horizontal dotted line shows the zero-range result of Tan's contact $C=0.118k_{\rm F}^4$ in the mean-field approximation~\cite{OHASHI2020103739}.}
    \label{fig:9}
\end{figure}
Figure~\ref{fig:9} shows the momentum distribution $n_{k}$ multiplied by $k^4$ to see its power-law behavior.
It is known that the high-momentum tail is given by Tan's contact $C$ as $\lim_{k\rightarrow\infty}n_{\bm{k}}\rightarrow\frac{C}{k^4}$ for the contact-type interaction.
Using the mean-field approximation, one can obtain $C=m^2|\Delta|^2=0.118k_{\rm F}^4$ at unitarity~\cite{OHASHI2020103739}. 
It can be obtained by expanding $E_{\bm{k}}=\sqrt{\{\xi_{k}+\Sigma(\bm{k})\}^2+|\Delta(\bm{k})|^2}$
with respect to $|\Delta|^2/\xi_{\bm{k}}^2$ in $n_{\bm{k}}$, where $\gamma_{k}=1$ for $r\rightarrow 0$.
Indeed, $k^4n_{k}$ at $k_{\rm F}r=0.01$ exhibits a plateau being consistent with $C=0.118k_{\rm F}^4$ (horizontal dotted line) in the region where $k_{\rm F}\lesssim k \lesssim \Lambda$.
If the density (or $k_{\rm F}r$) increases,
such a plateau cannot be visible because $\Delta(\bm{k})$ is suppressed in the high-momentum regime ($k\gesim\Lambda=\frac{2}{r}$) due to $\gamma_k$.
Therefore, while the nuclear contact formalism has been extensively discussed in nuclear systems by taking the momentum average~\cite{Hen2015PhysRevC.92.045205},
our result indicates that the dilute system with small $k_{\rm F}r$ is more suitable to address the contact parameter.
Considering the effective range of the isoscalar nucleon-nucleon interaction in the $^3S_1$ channel (i.e., $r_{^3S_1}=1.76$ fm~\cite{Wiringa1995PhysRevC.51.38}),
one can estimate that the relevant momentum region for original Tan's contact can be $k\lesssim 1.14$ fm$^{-1}$ in nuclear systems.
The higher momentum region can be affected by non-universal short-range sectors such as repulsive core.

\red{To see a more closed connection with the nuclear contact~\cite{WEISS2018211},
one may define a {\it nuclear-contact-like} quantity $\tilde{C}$, that is, the fraction of the single-particle density above $k=k_{\rm F}$ as
\begin{align}
\label{eq:nuclc}
    \tilde{C}=2\sum_{\bm{k}}\theta(k-k_{\rm F})n_{\bm{k}}.
\end{align}
Indeed, in the case of an ideal Fermi gas with $n_{\bm{k}}=\theta(k_{\rm F}-k)$, $\tilde{C}$ becomes exactly zero.
In this regard, $\tilde{C}$ can be regarded as a quantity characterizing the interaction effect.
If one assumes that the momentum distribution shows a power-law behavior as $n_{\bm{k}}\simeq \frac{C}{k^4}$ at $k\geq k_{\rm F}$, one can obtain
\begin{align}
\label{eq:nuclc2}
    \frac{\tilde{C}}{\rho}\simeq\frac{2}{(2\pi)^3\rho}\int_{k_{\rm F}}^{\infty}4\pi k^2dk\frac{C}{k^4}= \frac{3C}{k_{\rm F}^4},
\end{align}
indicating that $\tilde{C}$ is equivalent to $C$ except for the prefactor under this assumption.
}
\begin{figure}
    \centering
    \includegraphics[width=8.5cm]{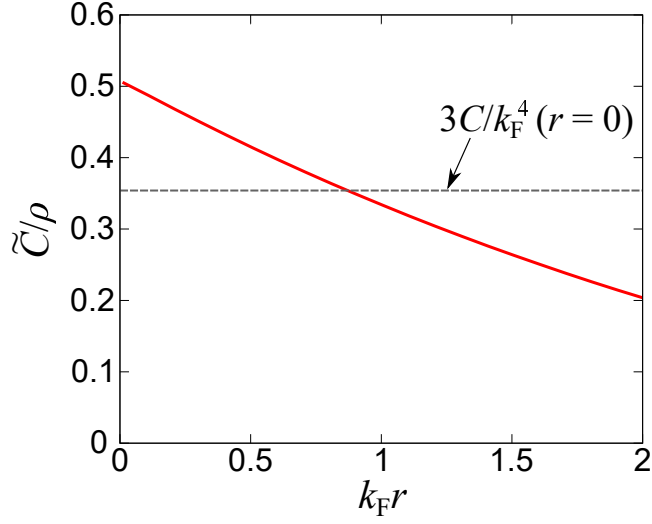}
    \caption{\red{Nuclear-contact-like parameter $\tilde{C}/\rho$ defined in Eq.~(\ref{eq:nuclc}) at $1/a=0$.
    The horizontal dashed line presents $3C/k_{\rm F}^4$ at the zero-range limit under the assumption that the distribution function is given by $n_{\bm{k}}\simeq\frac{C}{k^4}$ above $k=k_{\rm F}$ [see Eq.~(\ref{eq:nuclc2})].}}
    \label{fig:10}
\end{figure}
\red{Figure~\ref{fig:10} shows $\tilde{C}/\rho$ as a function of $k_{\rm F}r$ at $1/a=0$.
$\tilde{C}/\rho$ qualitatively manifests the fraction of interacting particles such as short-range correlated pairs.
It is found that $\tilde{C}/\rho$ monotonically decreases with increasing $k_{\rm F}r$.
On the one hand, such a behavior is consistent with $k_{\rm F}r$-dependence of $\Delta$ shown in Fig.~\ref{fig:3}(a).
On the other hand, $\tilde{C}/\rho$ quantitatively deviates from $3C/k_{\rm F}^4$ obtained in Eq.~(\ref{eq:nuclc2}) and accidentally $\tilde{C}/\rho\simeq 3C/k_{\rm F}^4$ around $k_{\rm F}r\simeq 1$.
This can be understood from the fact that
$n_{\bm{k}}$ is generally larger than $C/k^4$ around $k=k_{\rm F}$ as shown in Fig.~\ref{fig:9}.
However, $\tilde{C}/\rho$ becomes smaller than $3C/k_{\rm F}^4$ because the power-law behavior of $n_{\bm{k}}$ in the high-momentum regime is suppressed by the finite-range effect.
In this regard, one can find that $\tilde{C}$ characterizes the correlation effects on $n_{\bm{k}}$ near $k=k_{\rm F}$.
}

\section{Summary}
\label{sec:4}
In this study, we have theoretically investigated the role of the effective range $r$ in the density-induced BEC-BCS crossover.
The detailed structure of the interaction becomes crucial in the high-density regime of the density-induced BEC-BCS crossover in contrast to ultracold Fermi gases near the broad Feshbach resonance where the contact-type interaction characterized by the scattering length $a$ works well.

We have focused on the effective range correction which is a leading-order contribution towards non-universal short-range sectors in the $s$-wave phase shift.
Using the separable interaction potential exactly reproducing the phase shift within the effective-range expansion (i.e., without any higher-order coefficients such as shape parameter),
we have examined the superfluid properties in the density-induced BEC-BCS crossover.

Using the HFB theory, we have showed that the superfluid order parameter is strongly suppressed in the high-density regime due to the effective range corrections regardless of the sign of the scattering length.
In the high-density (large effective-range) limit, we have found that the interaction effect can be negligible because of the vanishing HF self-energy.
Although the mechanical collapse does not occur with the {\it pure} effective range correction,
the sound velocity exhibits the characteristic behavior along the density evolution.
While the sound velocity increases as the Bogoliubov phonon reflecting the repulsive molecule-molecule interaction in the low-density region ($k_{\rm F}r\lesssim 1$),
it turns to decrease in the crossover-density region ($1\lesssim k_{\rm F}r\lesssim \pi$) due to the Stoner enhancement of the compressibility.
Eventually, in the high-density regime ($k_{\rm F}r\gesim \pi$) where the zero-momentum Hartree approximation breaks down,
the sound velocity increases again towards the ideal-gas result $\frac{1}{\sqrt{3}}v_{\rm F}$ because the interaction effect is highly suppressed by the form factor and the HF self-energy shrinks in the momentum space.
\red{Moreover, the finite-range effect on the high-momentum tail of the distribution function and its relation to the nuclear contact has been discussed.}

For future work, it is important to include fluctuation effects beyond the mean-field approximation.
In particular, the density and spin fluctuations can be important in the high-density regime of neutron matter with realistic nucleon-nucleon interactions~\cite{ramanan2021pairing}.
Moreover, multi-body molecular interactions may appear in the BEC side~\cite{kagamihara2022isothermal}.
Also, it is interesting to investigate the mechanical collapse towards the electron-hole droplet as well as alpha condensates from uniform systems by using the finite-range interaction in the thermodynamic quartet BCS theory~\cite{Guo2022PhysRevC.105.024317,PhysRevResearch.4.023152}.
The relation with dense two-color QCD is also a fascinating topic~\cite{iida2022velocity}.
The low-dimensionality would be taken into account to consider the density-induced BEC-BCS crossover in recent superconducting materials~\cite{shi2021density}.
The non-separability of the interaction in the momentum space should be addressed in details in the future.
To investigate the relation with sound velocity peak in the hadron-quark crossover~\cite{kojo2021qcd}, one may consider the extension of such a study to three-body clustering crossover~\cite{Tajima2022PhysRevResearch.4.L012021}
and the case with additional repulsive interactions~\cite{Kojo2022PhysRevD.105.076001}.

\begin{acknowledgments}
The authors thank T. Hatsuda, K. Iida, E. Itou, T. Kojo, and H. Sakakibara for fruitful discussions.
H.T. acknowledges the JSPS Grants-in-Aid for Scientific Research under Grant Nos.~18H05406, ~22K13981.
H.L. acknowledges the JSPS Grant-in-Aid for Early-Career Scientists under Grant No.~18K13549, the JSPS Grant-in-Aid for Scientific Research (S) under Grant No.~20H05648, and the RIKEN Pioneering Project: Evolution of Matter in the Universe.
\end{acknowledgments}

\appendix

\section{Other separable potentials}
\label{app:A}
As another example,
we consider the separable interaction with the Gaussian form factor given by
\begin{align}
    U_{\rm G}(\bm{k},\bm{k}')=g_{\rm G}\gamma_{k,{\rm G}}\gamma_{k',{\rm G}},
\end{align}
where $\gamma_{k,{\rm G}}=e^{-(k/\Lambda_{\rm G})^2}$.
The two-body $T$-matrix reads
\begin{align}
\label{eq:tmatg}
    T_{\rm G}(\bm{k},\bm{k}';\omega)=\gamma_{k,{\rm G}}\left[\frac{1}{g_{\rm G}}-\Pi_{\rm G}(\omega)\right]^{-1}\gamma_{k',{\rm G}},
\end{align}
where
\begin{align}
    \label{eq:pi_g}
    \Pi_{\rm G}(\omega)&=\sum_{\bm{p}}\frac{\gamma_{p,{\rm G}}^2}{\omega_+-p^2/m}
\end{align}
is the two-particle propagator.
\red{The momentum summation can be replaced by the integration and performed analytically as}
\begin{align}
\label{eq:pi_g2}
    \Pi_{\rm G}(\omega)&=\red{-\frac{m}{4\pi^2}
    \left[
    \Lambda_{\rm G}\sqrt{\frac{\pi}{2}}+
    m\omega\int_{-\infty}^{\infty}dp\frac{e^{-2p^2/\Lambda_{\rm G}^2}}{p^2-m\omega_{+}}
    \right]}
    \cr
    &\red{=-\frac{m}{4\pi^2}
    \left[
    \Lambda_{\rm G}\sqrt{\frac{\pi}{2}}+
    i\pi\sqrt{m\omega}
    e^{-\frac{2m\omega}{\Lambda_{\rm G}^2}}\right]}\cr
    &\quad\red{+\frac{m\sqrt{m\omega}}{4\pi}
    e^{-\frac{2m\omega}{\Lambda_{\rm G}^2}}
    {\rm erfi}\left(\frac{\sqrt{2m\omega}}{\Lambda_{\rm G}}\right),}
\end{align}
\red{where we used}
\begin{align}
\label{eq:pi_g3}
    \red{\int_{-\infty}^{\infty}dp\frac{e^{-2p^2/\Lambda_{\rm G}^2}}{p^2-m\omega_{+}}
    =\frac{i\pi}{\sqrt{m\omega}}w\left(\frac{\sqrt{2m\omega}}{\Lambda_{\rm G}}\right)}
\end{align}
\red{In Eq.~(\ref{eq:pi_g3}),
$w(z)=\frac{2iz}{\pi}\int_{0}^{\infty}dx\frac{e^{-x^2}}{z^2-x^2}$ is the Faddeeva function,
which is associated with the imaginary error function ${\rm erfi}(z)$ as $w(z)=e^{-z^2}\left[1+i{\rm erfi}(z)\right]$. 
Using Eqs.~(\ref{eq:tmatg}) and (\ref{eq:pi_g2}), we obtain}
\begin{align}
\label{eq:psgauss}
    &-\frac{m}{4\pi}\left[-\frac{1}{a}+\frac{1}{2}rk^2-\mathcal{S}r^3k^4-ik\right]+O(k^5)\cr
    &\red{=e^{\frac{2k^2}{\Lambda_{\rm G}^2}}\left(\frac{1}{g_{\rm G}}+\frac{m\Lambda_{\rm G}}{4\pi^2}\sqrt{\frac{\pi}{2}}\right)
    +i\frac{mk}{4\pi}
    -\frac{mk}{4\pi}{\rm erfi}\left(\frac{\sqrt{2}k}{\Lambda_{\rm G}}\right).}
\end{align}
Thus, comparing the coefficients, we obtain
\begin{align}
    \frac{m}{4\pi a}=\frac{1}{g_{\rm G}}+\frac{m\Lambda_{\rm G}}{4\pi^2}\sqrt{\frac{\pi}{2}},
\end{align}
\begin{align}
    \red{-\frac{mr}{8\pi}=\frac{m}{2\pi a \Lambda_{\rm G}^2}
    -\frac{m}{\sqrt{2}\pi^{3/2}\Lambda_{\rm G}}
    }.
\end{align}
Using these equations, we can determine $g_{\rm G}$ and $\Lambda_{\rm G}$ from the values of $a$ and $r$ as
\begin{align}
    g_{\rm G}&=-\frac{4\pi}{m}\left[\frac{\Lambda_{\rm G}}{\sqrt{2\pi}}-\frac{1}{a}\right]^{-1},
\end{align}
\begin{align}
\label{eq:lambda_g}
    \red{\Lambda_{\rm G}=\frac{4}{\sqrt{2\pi}r}\left[1+\sqrt{1-\frac{\pi r}{2a}}\right],}
\end{align}
We note that the parameter with \red{$r/a<2/\pi$} is valid because $\Lambda_{\rm G}$ should be a real value. 
One may notice that the higher order terms such as the shape parameter $\mathcal{S}$ are nonzero in the left hand side of Eq.~(\ref{eq:psgauss}). 
Indeed, one can obtain
\begin{align}
    \red{\mathcal{S}=\frac{2}{r^3\Lambda_{\rm G}^3}\left(\frac{1}{a\Lambda_{\rm G}}-\frac{2}{3}\sqrt{\frac{2}{\pi}}\right).
    }
\end{align}
Also, in the large effective-range limit \red{at $1/a=0$}, we obtain
\begin{align}
\label{eq:ugrlarge}
    U_{\rm G}(k,k)\simeq-\frac{4\pi^2}{m\Lambda_{\rm G}}\sqrt{\frac{2}{\pi}}e^{-2k^2/\Lambda_{\rm G}^2}\rightarrow -\frac{4\pi^2}{m}\delta(k).
\end{align}
Eq.~(\ref{eq:ugrlarge}) is consistent with Eq.~(\ref{eq:Urlarge}) based on $\gamma_k$ in Eq.~(\ref{eq:gamma}).

We also examine Yamaguchi potential~\cite{Yamaguchi1954PhysRev.95.1628} given by
\begin{align}
    U_{\rm Y}(\bm{k},\bm{k}')=g_{\rm Y}\gamma_{k,{\rm Y}}\gamma_{k',{\rm Y}},
\end{align}
where
\begin{align}
    \gamma_{k,{\rm Y}}=\frac{1}{1+(k/\Lambda_{\rm Y})^2}.
\end{align}
In a same way with other separable potentials,
one can obtain the two-body $T$-matrix
\begin{align}
    T_{\rm Y}(\bm{k},\bm{k}';\omega)
    &=\gamma_{k,{\rm Y}}\left[\frac{1}{g_{\rm Y}}-\Pi_{\rm Y}(\omega)\right]^{-1}\gamma_{k',{\rm Y}},
\end{align}
where
\begin{align}
    \Pi_{\rm Y}(\omega)&=\sum_{\bm{p}}\frac{\gamma_{p,{\rm Y}}^2}{\omega_+-p^2/m}\cr
    &=\frac{m\Lambda_{\rm Y}^3}{8\pi}\frac{1}{(\sqrt{m\omega}+i\Lambda_{\rm Y})^2}.
\end{align}
In this regard, we obtain
\begin{align}
&-\frac{m}{4\pi}\left[-\frac{1}{a}+\frac{1}{2}rk^2-\mathcal{S}r^3k^4-ik\right]
    \cr
    &=\left[1+\left(\frac{k}{\Lambda_{\rm Y}}\right)^2\right]^2
    \left[\frac{1}{g_{\rm Y}}-\frac{m\Lambda_{\rm Y}^3}{8\pi}\frac{1}{(k+i\Lambda_{\rm Y})^2}\right].
\end{align}
From the comparison between both left and right hand sides, we obtain
\begin{align}
    \frac{m}{4\pi a}=\frac{1}{g_{\rm Y}}+\frac{m\Lambda_{\rm Y}}{8\pi},
\end{align}
\begin{align}
-\frac{mr}{8\pi}=\frac{m}{2\pi a \Lambda_Y^2}-\frac{3m}{8\pi\Lambda_Y}.
\end{align}
In this way, $g_{\rm Y}$ and $\Lambda_{\rm Y}$
can be determined as
\begin{align}
    g_{\rm Y}=-\frac{4\pi}{m}\left[\frac{\Lambda_{\rm Y}}{2}-\frac{1}{a}\right]^{-1},
\end{align}
\begin{align}
    \Lambda_{\rm Y}=\frac{3}{2r}\left[1+\sqrt{1-\frac{16r}{9a}}\right]^{-1},
\end{align}
which are consistent with Ref.~\cite{tajima2019superfluid}.
Moreover, we obtain the shape parameter
\begin{align}
    \mathcal{S}=\frac{4\pi}{mr^3}\frac{1}{g_{\rm Y}\Lambda_{\rm Y}^4}.
\end{align}
We note that the higher order coefficients [i.e., $O(k^5)$] in the phase shift are exactly zero in the case of the Yamaguchi potential.

\section{Comparison with screened Coulomb interaction}
\label{app:B}
\begin{figure}[t]
    \centering
    \includegraphics[width=8cm]{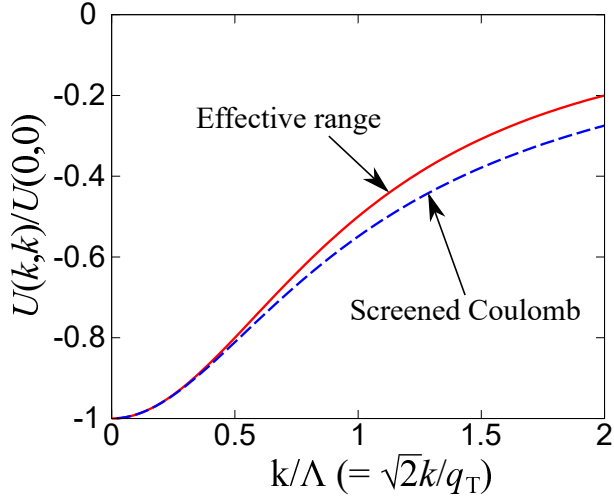}
    \caption{Comparison of the diagonal parts between the separable potential based on the effective range expansion and the screened Coulomb interaction.}
    \label{fig:11}
\end{figure}
We discuss the relation between the separable potential and the screened Coulomb interaction, which is given by
\begin{align}
    V_{\rm C}(\bm{q})&=-\frac{4\pi e^2}{q^2+q_{\rm T}^2}\equiv\red{-}\frac{4\pi e^2}{(\bm{k}-\bm{k}')^2+q_{\rm T}^2}
\end{align}
where $\bm{q}=\bm{k}-\bm{k}'$ is the relative momentum and $q_{\rm T}$ is the Thomas-Fermi wave-vector associated with the in-medium screening effect~\cite{fetter2012quantum}.
We use the partial wave decomposition
\begin{align}
    V_{\rm C}(\bm{k}-\bm{k}')=4\pi \sum_{\ell=0}U_{\ell}(k,k')
    \sum_{m=-\ell}^{m=\ell}Y_{m\ell}(\hat{\bm{k}})
    Y_{m\ell}^*(\hat{\bm{k}}'),
\end{align}
where $Y_{m\ell}^*(\hat{\bm{k}})$ is the spherical harmonics with $k=|\bm{k}|$ and $\hat{\bm{k}}=\bm{k}/k$.
Using $Y_{00}(\hat{\bm{k}})=1/\sqrt{4\pi}$,
we obtain the $s$-wave component
\begin{align}
    V_{\rm 0}(k,k')
    &=-\frac{\pi e^2}{kk'}\ln\left(\frac{k^2+k'^2+q_{\rm T}^2+2kk'}{k^2+k'^2+q_{\rm T}^2-2kk'}\right).
\end{align}
When we consider the diagonal part $k=k'$ in the momentum space, we obtain
\begin{align}
    V_0(k,k)&=-\frac{\pi e^2}{k^2}\ln\left(4\frac{k^2}{q_{\rm T}^2}+1\right)\cr
    &=-\frac{4\pi e^2}{q_{\rm T}^2}+\frac{8\pi e^2}{q_{\rm T}^4}k^2+O(k^4).
\end{align}
It can be associated with the separable potential
\begin{align}
    U(\bm{k},\bm{k})&=g\gamma_{k}^2\equiv g\frac{1}{1+(k/\Lambda)^2}\cr
    &=g\left[1-\left(\frac{k}{\Lambda}\right)^2+O(k^4)\right].
\end{align}
By comparing the coefficients, one obtains
\begin{align}
    g=-\frac{4\pi e^2}{q_{\rm T}^2},
\quad \Lambda=\frac{1}{\sqrt{2}}q_{\rm T}.
\end{align}
In such a case, the s-wave component of the screened Coulomb interaction can be rewritten as
\begin{align}
    V_0(k,k)=\frac{g\Lambda^2}{2k^2}\ln\left(2\frac{k^2}{\Lambda^2}+1\right).
\end{align}
Figure~\ref{fig:11} shows the comparison between the separable interaction for the effective range expansion and the screened Coulomb interaction for the diagonal component with respect to the relative momenta.
These two potentials agree well with each other in the low-momentum region $k\lesssim \Lambda\equiv\frac{1}{\sqrt{2}}q_{\rm T}$.
The deviation between two potentials can be
\begin{align}
    \delta V(k,k')= 3g\left(\frac{k}{\Lambda}\right)^4+O(k^6).
\end{align}
In this regard, our separable potential can be used to study low-energy pairing properties in electron-hole systems.
However, it is needed to evaluate $q_{\rm T}$ microscopically to investigate full density dependence~\cite{Conti2019PhysRevB.99.144517}.

\section{\red{Dilute BEC limit}}
\label{app:C}
\red{
In this appendix,
we derive the analytical expressions in the dilute BEC limit with $\mu<0$.
Here we neglect the HF self-energy $\Sigma(k)$ because $\Sigma(k)$ is negligible in the BEC regime as shown in Fig.~\ref{fig:4}(b). 
Assuming that $|\mu|$ is sufficiently large compared to $\Delta(k)\equiv\Delta\gamma_k$~\cite{OHASHI2020103739},
we expand the quasiparticle dispersion as
\begin{align}
\label{eq:C1}
    \frac{1}{E_{\bm{k}}}\simeq\frac{1}{2\xi_{\bm{k}}}-\frac{\Delta^2\gamma_k^2}{2\xi_{\bm{k}}^3}.
\end{align}
Using Eq.~(\ref{eq:C1}), we approximate Eqs.~(\ref{eq:gapeq}) and (\ref{eq:numeq}) as
\begin{align}
\label{eq:gapapp}
    0
    &\simeq
    \frac{m}{4\pi a}+
    \sum_{\bm{k}}\gamma_{k}^2\left[\frac{1}{2\xi_{\bm{k}}}-\frac{1}{2\varepsilon_{\bm{k}}}\right]
    -\Delta^2\sum_{\bm{k}}\frac{\gamma_k^4}{4\xi_{\bm{k}}^3},
\end{align}
\begin{align}
\label{eq:numapp}
        \rho&\simeq\Delta^2\sum_{\bm{k}}
    \frac{\gamma_k^2}{2\xi_{\bm{k}}^2}.
\end{align}
Eq.~(\ref{eq:gapapp}) without the third term is equivalent to the bound-state equation given by the pole of $T$-matrix as $[T(\bm{0},\bm{0},\omega=-E_{\rm b})]^{-1}=0$ [see also Eq.~(\ref{eq:tmatrix})], when one takes $2|\mu|=E_{\rm b}$ in Eq.~(\ref{eq:gapapp}). In this regard, one can find that $|\mu|\simeq E_{\rm b}/2$ is realized in the dilute BEC limit with $a>0$.
The momentum summation in Eqs.~(\ref{eq:gapapp}) and (\ref{eq:numapp}) can be performed analytically as
\begin{align}
\label{eq:int1}
    \sum_{\bm{k}}\gamma_{k}^2\left[\frac{1}{2\xi_{\bm{k}}}-\frac{1}{2\varepsilon_{\bm{k}}}\right]
    &=\frac{m\Lambda^2}{4\pi(\sqrt{2m|\mu|}+\Lambda)}-\frac{m\Lambda}{4\pi},
\end{align}
\begin{align}
\label{eq:int2}
    \sum_{\bm{k}}\frac{\gamma_k^4}{4\xi_{k}^3}
    &=
    \frac{m^3\Lambda^3(4\sqrt{2m|\mu|}+\Lambda)}{16\pi(2m|\mu|)^{3/2}(\sqrt{2m|\mu|}+\Lambda)^4},
\end{align}
\begin{align}
\label{eq:int3}
    \sum_{\bm{k}}\frac{\gamma_k^2}{2\xi_{\bm{k}}^2}
    &=\frac{m^2\Lambda^2}{4\pi \sqrt{2m|\mu|}(\sqrt{2m|\mu|}+\Lambda)^2}.
\end{align}
Using Eqs.~(\ref{eq:gapapp}), (\ref{eq:int1}) and (\ref{eq:int2}), we obtain
\begin{align}
\label{eq:c7}
    |\mu|=\frac{E_{\rm b}}{2}-\Delta^2
    \frac{ma\Lambda^3(4\sqrt{mE_{\rm b}}+\Lambda)}{4\sqrt{mE_{\rm b}}(\sqrt{mE_{\rm b}}+\Lambda)^4}+O(\Delta^4/E_{\rm b}^4).
\end{align}
Also, Eqs.~(\ref{eq:numapp}) and (\ref{eq:int3}) lead to 
\begin{align}
\label{eq:c8}
    \Delta^2&=\frac{4\pi\rho\sqrt{2m|\mu|}}{m^2}\left(1+\frac{\sqrt{2m|\mu|}}{\Lambda}\right)^2\cr
    &\equiv\frac{16}{3\pi}E_{\rm F}^{3/2}|\mu|^{1/2}
    \left(1+\frac{\sqrt{2m|\mu|}}{\Lambda}\right)^2.
\end{align}
Because we find $|\mu|\simeq E_{\rm b}/2$ at the leading order of $\Delta/E_{\rm b}$,
we obtain Eq.~(\ref{eq:gapBEC}) by substituting $|\mu|\simeq E_{\rm b}/2$ into Eq.~(\ref{eq:c8}).
The resulting $\Delta$ is consistent with the previous work (e.g., Ref.~\cite{OHASHI2020103739,STRINATI20181}) in the zero-range limit (or $\Lambda\rightarrow \infty$).
}

\red{It is known that the second term of Eq.~(\ref{eq:c7}) is associated with the molecule-molecule repulsion~\cite{OHASHI2020103739}.
This fact originates from the equivalence between the gap equation and the Hugenholtz-Pines (HP) condition of molecular bosons within the mean-field level in the BEC limit~\cite{PhysRevLett.91.030401,PhysRevB.98.104507}.
It is useful to introduce the molecular chemical potential $\mu_{\rm B}=2\mu+E_{\rm b}$.
Using Eq.~(\ref{eq:c7}), we obtain
\begin{align}
    \mu_{\rm B}&=\rho_{\rm B}\frac{4\pi}{m_{\rm B}}
    \frac{2a\Lambda(4\sqrt{mE_{\rm b}}+\Lambda)}{(\sqrt{mE_{\rm b}}+\Lambda)^2},
\end{align}
where $\rho_{\rm B}=\rho/2$ and $m_{\rm B}=2m$ are the molecular density and mass, respectively. 
Considering the HP condition, one can relate $\mu_{\rm B}=U_{\rm BB}\rho_{\rm B}$ (where $U_{\rm BB}$ is the coupling constant of molecule-molecule repulsion) and the molecule-molecule scattering length $a_{\rm BB}$ as shown in Eq.~(\ref{eq:abb}).
}

\bibliographystyle{apsrev4-2}
\bibliography{reference.bib}

%\begin{thebibliography}{99}
%\bibitem{Sheehy} K. R. Patton and D. E. Sheehy,
%Phys. Rev. A \textbf{83}, 051607(R) (2011).
%\end{thebibliography}
%\end{CJK}
\end{document}